\documentclass[aps, twocolumn, a4paper, prb, superscriptaddress]{revtex4-1}
\usepackage[utf8]{inputenc}
\usepackage{amsmath}
\usepackage{amsfonts}
\usepackage{amssymb}
\usepackage{graphicx}
\usepackage{hyperref}
\usepackage{siunitx}
\hypersetup{citecolor=blue,filecolor=blue,linkcolor=blue,menucolor=blue,urlcolor=black,colorlinks=true}

\newcommand{\NTT}{\affiliation{NTT Basic Research Laboratories, NTT Corporation, 3-1 Morinosato-Wakamiya, Atsugi, Kanagawa, 243-0198, Japan}}
\newcommand{\TUS}{\affiliation{Department of Applied Physics, Faculty of Science, Tokyo University of Science, 6-3-1 Niijuku, Katsushika, Tokyo, 125-8585, Japan}}
\newcommand{\AIST}{\altaffiliation[Present Address: ]{Nanoelectronics Research Institute, National Institute of Advanced Industrial Science and Technology, 1-1-1 Umezono, Tsukuba, Ibaraki, 305-8568, Japan}}

\begin{document}
\author{Hiraku~Toida}
\email{hiraku.toida.ds@hco.ntt.co.jp}
\NTT
\author{Takuya~Ohrai}
\NTT
\TUS
\author{Yuichiro~Matsuzaki}
\AIST
\NTT
\author{Kosuke~Kakuyanagi}
\NTT
\author{Shiro~Saito}
\NTT
\TUS

\title{Control of transition frequency of a superconducting flux qubit by longitudinal coupling to the photon number degree of freedom in a resonator}

\begin{abstract}
We control transition frequency of a superconducting flux qubit coupled to a frequency-tunable resonator comprising a direct current superconducting quantum interference device (dc-SQUID) by microwave driving.
The dc-SQUID mediates the coupling between microwave photons in the resonator and a flux qubit.
The polarity of the frequency shift depends on the sign of the flux bias for the qubit and can be both positive and negative.
The absolute value of the frequency shift becomes larger by increasing the photon number in the resonator.
These behaviors are reproduced by a model considering the magnetic interaction between the flux qubit and dc-SQUID.
The tuning range of the transition frequency of the flux qubit reaches $\approx$\SI{1.9}{GHz}, which is much larger than the ac Stark/Lamb shift observed in the dispersive regime using typical circuit quantum electrodynamics devices.
\end{abstract}

\maketitle
Implementing a large-scale quantum system requires controllable qubits with excellent coherence properties.
A superconducting qubit is one of the most promising candidates for implementing such a system with solid-state devices \cite{Arute2019, Corcoles2019, Otterbach2017, Gong2019}.
The transition frequency of a superconducting qubit is commonly controlled by applying magnetic flux to the superconducting loop of a direct current superconducting quantum interference device (dc-SQUID) \cite{Nakamura1999} or a flux qubit \cite{Mooij1999}.
However, this standard method requires at least one wire to control one qubit, and the wiring of a large-scale system would be technically challenging
in terms of packing and reducing crosstalk.
Furthermore, it cannot be applied for pulse control of a qubit in a 3D cavity \cite{Paik2011, Rigetti2012, Stern2014, Reagor2016, Abdurakhimov2019} without adding wiring to the cavity, 
which weakens our ability to engineer the electromagnetic environment provided by the cavity.

An alternative method to control a superconducting qubit is to use a circuit quantum electrodynamics (QED) architecture \cite{Blais2004, Wallraff2004}.
With this architecture, the transition frequency of a qubit can be controlled by microwave driving with an ac Stark or a Lamb shift in a dispersive regime.
In such a case, the tuning range of the transition frequency is limited to the order of \SI{100}{MHz} because of the dispersive condition \cite{Schuster2005, Schuster2007a, Schuster2007, Ong2011}.

In general, there are two types of qubit-resonator couplings: transversal and longitudinal \cite{Billangeon2015a}.
The circuit QED architecture is a prime example of transversal coupling of a qubit to the displacement degree of freedom in a resonator.
On the other hand, longitudinal coupling has recently been the focus of research for fast qubit readout \cite{Billangeon2015a, Didier2015, Touzard2019, Ikonen2019} or coupling between two qubits \cite{Billangeon2015a, Royer2017, Geller2015a, Weber2017a}.

Here, we present an alternative method in which the transition frequency of a flux qubit is controlled through its longitudinal coupling to the photon number degree of freedom in a resonator.
A frequency-tunable resonator comprising a dc-SQUID and capacitors works as a mediator between microwave photons in the resonator and the magnetic flux through the flux qubit because of the inductive coupling between the dc-SQUID and qubit.
Due to the longitudinal magnetic coupling, the flux qubit experiences an effective magnetic flux generated by the microwave photons in the resonator, and thus the transition frequency of the flux qubit can be controlled.
By increasing the number of microwave photons in the resonator, the transition frequency of the flux qubit is successfully controlled up to $\approx$ \SI{1.9}{GHz}.

\begin{figure}
\includegraphics{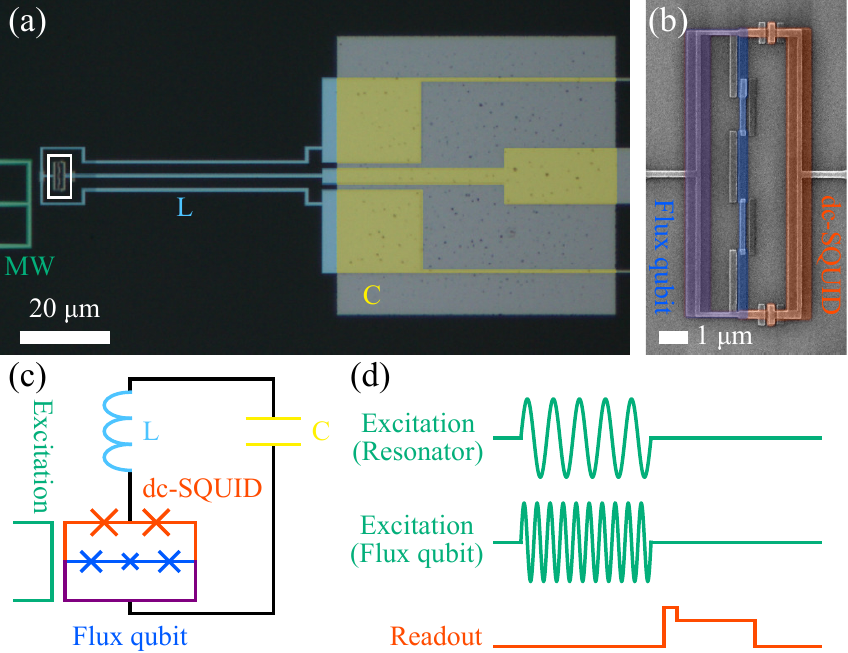}
\caption{(a) False color optical microscope image of the device. Microwave (MW) line (green), wire inductor L (sky blue), and parallel plate capacitor C (yellow).
(b) False color scanning electron microscope image of the dc-SQUID and flux qubit [white box in (a)]. The dc-SQUID (red) and flux qubit (blue) share the edges of the loops (purple).
(c) Equivalent circuit model of the device. The color scheme is the same as that in (a) and (b). The readout circuit for the dc-SQUID is omitted.
(d) Pulse sequence for the spectroscopy of the dc-SQUID and/or flux qubit. Excitation signals are applied to the MW line, and a qubit readout pulse is sent to the dc-SQUID.}
\label{fig:1}
\end{figure}

Figure~\ref{fig:1}(a) and (b) respectively show an optical microscope image of the fabricated device and a scanning electron microscope image of the superconducting flux qubit \cite{Mooij1999, Orlando1999} coupled to a frequency-tunable resonator containing the dc-SQUID \cite{Johansson2006}.
The lumped-element resonator consists of parallel plate capacitors (C), line inductors (L), and the dc-SQUID [Fig.~\ref{fig:1}(c)] \cite{Mutus2013}.
The flux qubit and the dc-SQUID have shared edges \cite{Chiorescu2003}, which ensures that the inductive coupling between them is strong enough.
To excite the tunable resonator and flux qubit, we radiate the microwave pulse shown in Fig.~\ref{fig:1}(d) to them through the same on-chip microwave line (MW).
The resonator and the qubit states are read out by the switching probability of the dc-SQUID \cite{Deppe2007}.
The operating point of the tunable resonator and the flux qubit is controlled by applying an external magnetic field with a superconducting magnet.
All the experiments were performed in a dilution refrigerator with a  base temperature of about \SI{25}{mK}.

\begin{figure}
\centering
\includegraphics{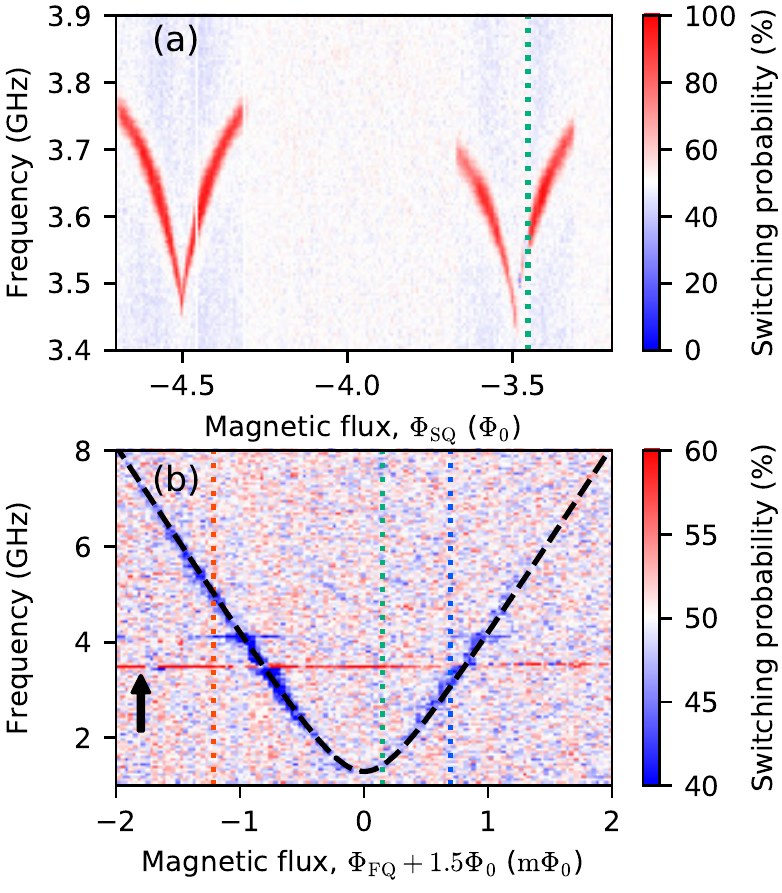}
\caption{(a) Spectrum of the frequency-tunable resonator as a function of applied flux $\Phi_{\mathrm{SQ}}$ to the dc-SQUID. The operating point of the flux qubit is indicated by the green dotted line. 
(b) Spectrum of the flux qubit as a function of applied flux $\Phi_{\mathrm{FQ}}$ to the flux qubit. The resonance line indicated by the black arrow originates from the frequency-tunable resonator. Operating points used in two-tone spectroscopy are indicated by red, green, and blue dotted lines. The dashed line is the fit to the model [Eq. (\ref{eq:FQHamiltonian})].}
\label{fig:2}
\end{figure}

First, the properties of the tunable resonator and flux qubit are characterized independently.
Figure~\ref{fig:2}(a) shows the spectrum of the tunable resonator as a function of the magnetic flux through the dc-SQUID loop $\Phi_{\mathrm{SQ}}$.
The resonance angular frequency of the tunable resonator, $\omega_\mathrm{r}$, can be controlled by $\Phi_{\mathrm{SQ}}$:
\begin{equation}
    \omega_\mathrm{r}\left(\Phi_{\mathrm{SQ}}\right) = \omega_\mathrm{LC}\frac{1}{\sqrt{1+L_\mathrm{SQ}\left(\Phi_{\mathrm{SQ}}\right)/L}},
    \label{eq:Resonator}
\end{equation}
where $\omega_{\mathrm{LC}} = \left(LC\right)^{-1/2}$ is the resonance angular frequency of the LC resonator without the dc-SQUID, and $L_\mathrm{SQ}$ is the effective inductance of the dc-SQUID controlled by $\Phi_\mathrm{SQ}$.
The spectrum is missing around $\Phi_\mathrm{SQ} \approx -4.0 \Phi_0$, possibly because it is affected by an unwanted resonance around \SI{4}{GHz}.

Figure~\ref{fig:2}(b) shows the spectrum of the flux qubit as a function of applied magnetic flux $\Phi_{\mathrm{FQ}}$ to the flux qubit loop.
The spectrum is reproduced by calculating the eigenenergy of the following Hamiltonian \cite{Wal2000}:
\begin{equation}
    H_\mathrm{FQ} = \frac{\Delta}{2}\sigma_x + \frac{\varepsilon(\Phi_{\mathrm{FQ}})}{2}\sigma_z,
    \label{eq:FQHamiltonian}
\end{equation}
where $\sigma_i$ $(i = x, z)$ is the Pauli operator, $\Delta$ is the energy gap, $\varepsilon(\Phi_{\mathrm{FQ}}):=2I_\mathrm{p}\left[\Phi_{\mathrm{FQ}}-\left(n/2\right)\Phi_0\right]$ is the energy detuning with $n$ being 
an odd integer, $I_\mathrm{p}$ is the persistent current, $\Phi_0:=h/2e$ is the magnetic flux quanta, $h$ is the Planck's constant, and $e$ is the elementary charge.
Here, $n = -3$ is selected \cite{Zhu2010}.
From the fitting to the flux qubit spectrum, the energy gap $\Delta/h$ is estimated to be \SI{1.30}{GHz}.
The persistent current is also extracted to be $I_\mathrm{p}\approx$ \SI{640}{nA} from the slope of the spectrum.
In the flux qubit spectrum, the horizontal straight line around \SI{3.45}{GHz} indicated by the black arrow is the resonance of the tunable resonator.
It can be safely assumed that the frequency of the tunable resonator is almost constant in the qubit spectrum, because the flux range to tune the qubit is much narrower than that for tuning the resonator.
However, it is important to note that the gradient of the tunable resonator spectrum is finite at the magnetic flux of the qubit operating point indicated by the green dotted line in Fig.~\ref{fig:2}(a).
This is necessary for this scheme because the interaction between the resonator and qubit occurs because of the flux coupling between them.
If the slope is finite, the resonator's frequency is controlled by the magnetic flux generated by the qubit, and vice versa.

\begin{figure*}
\centering
\includegraphics{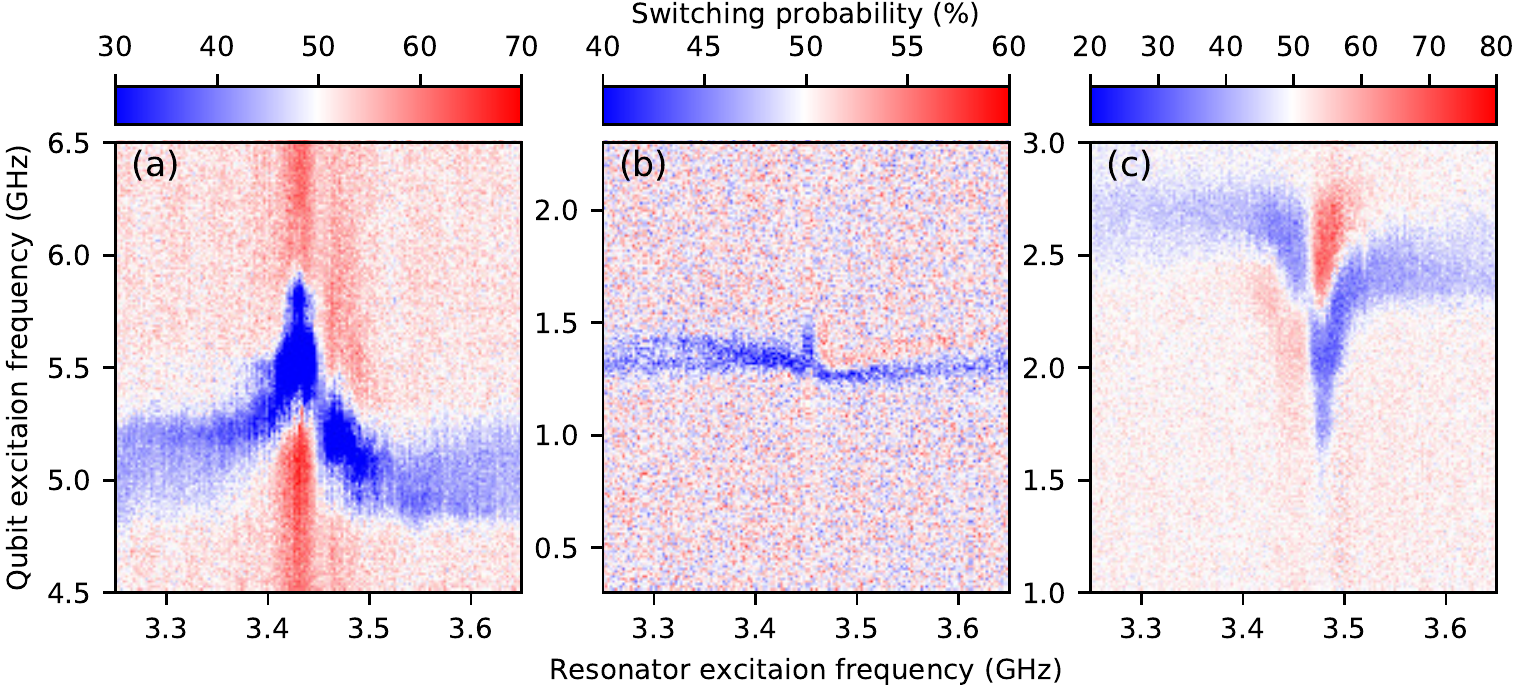}
\caption{Two-tone spectroscopy. Flux bias for the flux qubit is (a) $-1.05$, (b) $0.13$, or (c) $0.60$ m$\Phi_0$, respectively. 
These flux biases are indicated by the dotted lines in Fig.~\ref{fig:2}(b). Excitation power to the flux qubit and the resonator is the same for (a), (b), and (c).}
\label{fig:3}
\end{figure*}

Next, two-tone spectroscopy was performed to control the transition frequency of the flux qubit through the excitation to the frequency-tunable resonator. 
In addition to a microwave tone for the qubit excitation (0.25 to \SI{6.5}{GHz}), a secondary tone was applied to excite the frequency-tunable resonator (3.2 to \SI{3.7}{GHz}).
Figures~\ref{fig:3} show the results of the two-tone spectroscopy.
For this experiment, the operating point of the flux qubit was fixed at either $-1.05$, $0.13$, or $0.60$ m$\Phi_0$ indicated by red, green, and blue dotted lines, respectively in Fig.~\ref{fig:2}(b).
For these three experiments, the microwave excitation power to the resonator and flux qubit was fixed.
If the flux bias is negative, the transition frequency of the flux qubit increases when the resonator is excited around \SI{3.48}{GHz} [Fig.~\ref{fig:3}(a)].
On the other hand, if the flux bias is positive, the transition frequency of the flux qubit decreases [Fig.~\ref{fig:3}(c)].
It is also confirmed that the transition frequency changes little if near-zero flux bias is applied to the flux qubit [Fig.~\ref{fig:3}(b)].

\begin{figure}
\includegraphics{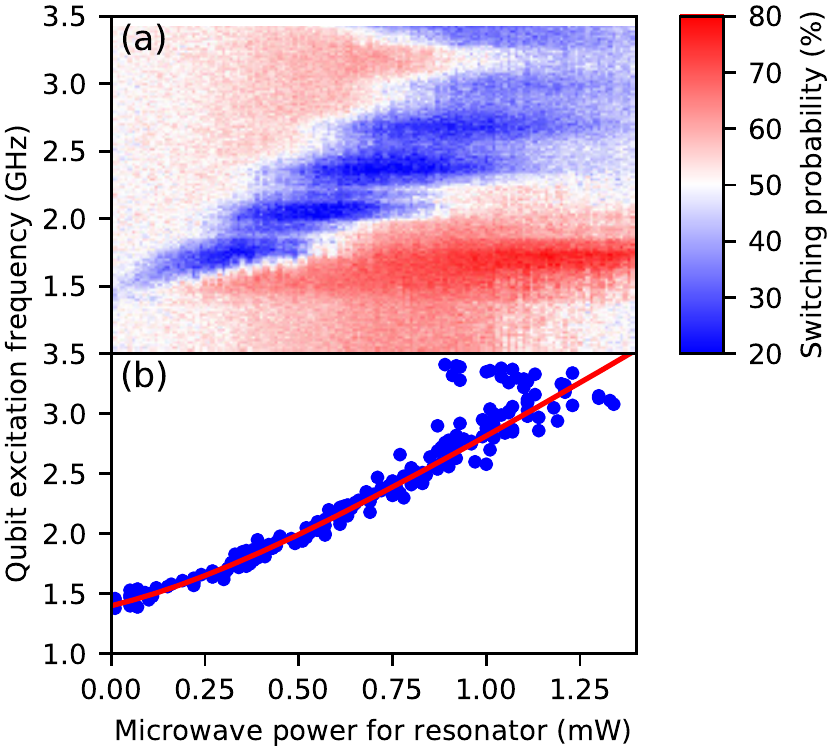}
\caption{Resonator excitation power dependence of the transition frequency of the flux qubit. The excitation frequency for the tunable resonator is fixed on resonance. 
(a) Qubit spectrum as a function of the excitation power of the resonator. The operation point of the flux qubit is fixed at $-0.067$ m$\Phi_0$. The overall structure has a small ripple along the vertical axis due to the presence of parasitic resonances in the measurement setup. 
(b) Transition frequencies of the flux qubit extracted from (a). The solid line is the fit to the model [Eq. (\ref{eq:freqshift})] using the data below \SI{0.8}{mW}. Microwave power for the resonator was measured at the microwave generator.}
\label{fig:4}
\end{figure}

To investigate the frequency shift in more detail, the flux qubit spectrum was measured as a function of the excitation power to the resonator as shown in Fig.~\ref{fig:4}(a).
For this experiment, the flux bias for the flux qubit was set to almost zero ($-0.067$ m$\Phi_0$), and the microwave tone for the resonator excitation was fixed on resonance.
As shown in Fig.~\ref{fig:4}(b), the transition frequency of the flux qubit increases linearly if the excitation power is large enough.
It is important to note that the transition frequency converges to the energy gap of the flux qubit, $\Delta$, with decreasing excitation power.

These experimental observations are explained by the total Hamiltonian of the system:
\begin{align}
    \label{eq:TotalHamiltonian}
    H &= H_{\mathrm{FQ}} + H_{\mathrm{r}}, \\
    \label{eq:ResonatorHamiltonian}
    H_{\mathrm{r}} &= \hbar \omega_{\mathrm{r}}(\Phi_{\mathrm{SQ}}) \left(a^\dagger a + \frac{1}{2}\right).
\end{align}
Since there is a magnetic interaction between the dc-SQUID and flux qubit, the resonance frequency of the tunable resonator is controlled by the state of the flux qubit.
The persistent current of the flux qubit, $I_\mathrm{p}$, generates the magnetic flux through the dc-SQUID, $MI_\mathrm{p}$, where $M$ is the mutual inductance between the dc-SQUID and flux qubit.
The resonance frequency of the tunable resonator is approximated by the following formula:
\begin{equation}
    \omega_{\mathrm{r}}(\Phi_\mathrm{SQ}) \approx \omega_{\mathrm{r}}^{0}(\Phi_\mathrm{SQ}) - \frac{d \omega_{\mathrm{r}}}{d \Phi_{\mathrm{SQ}}}MI_\mathrm{p} \sigma_z,
    \label{eq:Omega_r}
\end{equation}
where $\omega_{\mathrm{r}}^{0}$ is the bare resonator frequency without perturbation from the flux qubit.
Here, $\sigma_z$ expresses the direction of the circulating current of the flux qubit.
The second term of Eq. (\ref{eq:Omega_r}) has a negative sign because the operation point of the flux qubit is near $n=-3$.
From Eqs. (\ref{eq:FQHamiltonian}), (\ref{eq:TotalHamiltonian}), (\ref{eq:ResonatorHamiltonian}), and (\ref{eq:Omega_r}), the total Hamiltonian of the system is derived as follows:
\begin{equation}
    H = \frac{\Delta}{2}\sigma_x + \left[\frac{\varepsilon}{2} - \frac{g}{2} \left(a^\dagger a + \frac{1}{2}\right)\right] \sigma_z + \hbar \omega_{\mathrm{r}}^{0} \left(a^\dagger a + \frac{1}{2}\right),
\label{eq:SystemHamiltonian}
\end{equation}
where $g := 2\hbar\left( d\omega_{\mathrm{r}}/d\Phi_{\mathrm{SQ}}\right)MI_{\mathrm{p}} $ is the coupling strength between the frequency-tunable resonator and flux qubit.
From the eigenvalue of the Hamiltonian, the transition frequency of the flux qubit $f_{\mathrm{FQ}}$ is expressed as follows:
\begin{equation}
    hf_{\mathrm{FQ}} = \sqrt{\left[\varepsilon-g \left(N+\frac{1}{2}\right) \right]^2+\Delta^2},
    \label{eq:freqshift}
\end{equation}
where $N$ is the photon number in the resonator.
This expression is linearized if the condition $\left|\varepsilon-g\left(N+1/2\right)\right| \gg \Delta$ is satisfied:
\begin{equation}
    hf_{\mathrm{FQ}} \approx \left|\varepsilon-g\left(N+\frac{1}{2}\right)\right|.
    \label{eq:freqshift_lin}
\end{equation}

The model quantitatively explains the experimental results.
In addition to energy detuning $\varepsilon$, the model has additional tunability of $f_{\mathrm{FQ}}$ stemming from the $gN$ term.
From Eq. (\ref{eq:freqshift_lin}), we can explain the dependence of the polarity of the shift of $f_{\mathrm{FQ}}$ on $\varepsilon$.
If $\varepsilon$ is negative [positive], $f_{\mathrm{FQ}}$ increases [decreases] as the photon number increases, which is observed in Fig.~\ref{fig:3}(a) [(c)].
To understand the phenomenon observed in Fig.~\ref{fig:3}(b), the equation before linearization [Eq. (\ref{eq:freqshift})] should be used, because the effect of the energy gap $\Delta$ cannot be ignored.
In this case, the effect of the microwave photons in the resonator is not large compared to the cases of Figs.~\ref{fig:3}(a) and (c), which is consistent with the model.

Next, we investigate the shift of $f_{\mathrm{FQ}}$ as a function of the excitation power to the resonator [Fig.~\ref{fig:4}(b)].
Deviation from the linear trend is also observed in the low-power regime.
This behavior is interpreted as the effect of the energy gap $\Delta$ as explained by Eq. (\ref{eq:freqshift}).

Now, the coupling strength between the flux qubit and the dc-SQUID $g$ is estimated.
From the device design parameters and individual experimental results for the flux qubit and resonator, the coupling strength is derived using the relationship 
$g=2\hbar \left(d\omega_{\mathrm{r}}/d\Phi_{\mathrm{SQ}}\right)MI_\mathrm{p}$.
Here, the mutual inductance between the flux qubit and dc-SQUID, $M\approx$ \SI{12.1}{pH}, is estimated by numerical simulation using FastHenry \cite{Kamon1993}.
The persistent current of the flux qubit, $I_\mathrm{p}$, is derived from the flux qubit spectrum [Fig.~\ref{fig:2}(b)] as previously shown.
The slope of the resonator spectrum, $d\omega_{\mathrm{r}}/d\Phi_{\mathrm{SQ}}\approx$ 2$\pi\times$2.1 GHz/$\Phi_0$, is directly derived from Fig.~\ref{fig:2}(a).
By combining these values, the coupling strength is estimated to be $g\approx h\times$\SI{15.6}{MHz}.

It is important to emphasize the difference between the scheme presented here and a similar interaction of dispersive coupling between a qubit and resonator.
In the circuit QED experiments in dispersive regime $\delta \gg g_\mathrm{c}$ [$\delta$ $(g_\mathrm{c})$ is the detuning (coupling strength) between the resonator and the qubit], the interaction
Hamiltonian is approximated as $\left(g_\mathrm{c}^2/\delta\right)a^{\dagger }a\sigma _z $, and an ac Stark/Lamb shift is observed when we drive the resonator \cite{Blais2004}.
However, in the case of dispersive coupling, the qubit frequency shift is relatively small because a large detuning $\delta$ suppresses the shift as $g_\mathrm{c}^2/\delta \ll g_c$.
Moreover, if we drive the resonator too strongly, the number of photons increases, which results in the violation of the dispersive approximation.
On the other hand, there is no detuning dependence of the qubit frequency shift in our system.
Our method also has the advantage that the increase in the number of photons does not change the form of the Hamiltonian in the Eq. (\ref{eq:SystemHamiltonian}), although the frequency shift is technically limited by the critical current of the dc-SQUID.
For these reasons, the qubit frequency can be controlled in a broad range without fundamental limitations.
The sample used in the experiment showed the maximum frequency-tuning range of \SI{1.9}{GHz}.
This value is much larger than the typical case of circuit QED experiments in the order of $\approx$ \SI{100}{MHz} \cite{Schuster2005, Schuster2007a, Schuster2007, Ong2011}.

In conclusion, by coupling a frequency-tunable resonator with a flux qubit, we demonstrated frequency control of the flux qubit, where the shift increases as the number of photons increases.
Depending on the operation point of the flux qubit, either a positive or negative frequency shift is observed.
The tuning range of the qubit frequency reaches \SI{1.9}{GHz}.
A model using longitudinal magnetic coupling between the flux qubit and frequency-tunable resonator quantitatively
explains the experimental results with the coupling constant on the order of \SI{10}{MHz}.
Our method to control a flux qubit would be useful in implementing a large-scale quantum circuit with a smaller number of control lines or could provide further tunability to a flux qubit in a 3D cavity \cite{Abdurakhimov2019} without adding galvanic wiring into it.

We thank Mao-Chuang~Yeh, Anthony~J.~Leggett, and Hiroshi~Yamaguchi for helpful discussions.
This work was supported in part by MEXT Grant-in-Aid for Scientific Research on Innovative Areas ``Science of hybrid quantum systems'' (Grant No. 15H05867).

%\bibliographystyle{apsrev4-1}
%\bibliography{FQ_SQUID}


\end{document}